\pdfoutput=1
\documentclass[11pt,a4paper]{article}

\usepackage[T1]{fontenc}
\usepackage[utf8]{inputenc}
\usepackage[sc]{mathpazo}
\usepackage[scaled=0.86]{helvet}
\usepackage{microtype}

\usepackage[a4paper,top=2.6cm,bottom=2.8cm,left=2.7cm,right=2.7cm,headsep=0.8cm]{geometry}
\usepackage{amsmath,amssymb}
\usepackage{graphicx}
\usepackage{booktabs}
\usepackage[dvipsnames]{xcolor}
\definecolor{accent}{HTML}{000000}

\usepackage{titlesec}
\titleformat{\section}{\sffamily\Large\bfseries\color{accent}}{\thesection}{0.8em}{}
\titleformat{\subsection}{\sffamily\large\bfseries\color{accent}}{\thesubsection}{0.8em}{}
\titleformat{\paragraph}[runin]{\normalfont\itshape}{}{0em}{}[.]
\titlespacing*{\section}{0pt}{2.2ex plus 0.5ex minus 0.2ex}{1.0ex}
\titlespacing*{\subsection}{0pt}{1.8ex plus 0.4ex minus 0.2ex}{0.7ex}
\titlespacing*{\paragraph}{0pt}{1.2ex plus 0.3ex}{0.6em}

\usepackage{fancyhdr}
\fancypagestyle{plain}{\fancyhf{}\fancyfoot[C]{\sffamily\footnotesize\thepage}}

\usepackage[section]{placeins}
\usepackage[font=small,labelfont={bf,sf,color=accent},labelsep=period,skip=6pt,margin=0.5em]{caption}

\usepackage[authoryear,round]{natbib}
\usepackage{url}
\usepackage[colorlinks=true,linkcolor=accent,citecolor=accent,urlcolor=accent,
            pdftitle={Same mortality change, different life expectancy change},
            pdfauthor={Markus Sauerberg},
            pdfkeywords={life expectancy, counterfactual changes, decomposition, formal demography}]{hyperref}

\newcommand{\dd}{\,\mathrm{d}}

\makeatletter
\renewcommand{\maketitle}{%
  \thispagestyle{plain}%
  \begin{center}
    {\sffamily\color{accent}\rule{\textwidth}{2.5pt}}\par\vspace{1.0em}
    {\sffamily\LARGE\bfseries\color{accent}Same mortality change, different life expectancy change\par}
    \vspace{0.5em}
    {\sffamily\Large Counterfactual changes in life expectancy and the analogy to age-standardized death rates\par}
    \vspace{1.0em}
    {\sffamily\color{accent}\rule{\textwidth}{1pt}}\par\vspace{1.2em}
    {\large Markus Sauerberg\par}
    \vspace{0.3em}
    {\small Hamburg Cancer Registry, Hamburg, Germany\\
     \href{mailto:sauerbergmarkus@gmail.com}{\nolinkurl{sauerbergmarkus@gmail.com}}\par}
    \vspace{0.6em}
    {\small November 2026\par}
  \end{center}
  \vspace{0.8em}}
\renewenvironment{abstract}{%
  \begin{center}\begin{minipage}{0.88\textwidth}
  \noindent{\sffamily\bfseries\color{accent}Abstract}\par\vspace{0.4em}\small}
  {\end{minipage}\end{center}\vspace{0.8em}}
\makeatother

\title{Same mortality change, different life expectancy change: Counterfactual changes in life expectancy and the analogy to age-standardized death rates}
\author{Markus Sauerberg}
\date{November 2026}

\begin{document}
\maketitle

\begin{abstract}
\noindent
Changes in life expectancy at birth are widely used to compare the development of population health between countries, regions or sexes, and differences in these changes are usually attributed to differences in the development of mortality. However, the same relative change in age-specific mortality rates leads to different changes in life expectancy at birth, depending on the mortality schedule from which a population starts. This is because changes in life expectancy at birth can be expressed as the weighted sum of relative changes in age-specific mortality, with weights reflecting the initial mortality schedule. This structure is similar to that of the crude death rate, in which age-specific mortality rates are weighted by the population age structure. Using this analogy, I introduce counterfactual changes in life expectancy at birth, which evaluate the observed changes in mortality with the weights of a common reference population, and propose a decomposition of the difference between two changes in life expectancy into a mortality effect and a sensitivity effect, analogous to Kitagawa's decomposition of crude death rates. Using data from the Human Mortality Database, I apply both measures in three empirical applications. Between 1990 and 2019, the larger gains in life expectancy at birth of men compared with women were to a considerable extent a sensitivity effect, i.e., about two thirds in Germany and the entire difference in Spain and Poland. Further, I find that almost a quarter of the larger gain of East German compared with West German men reflects their higher initial mortality. Finally, for life expectancy changes between 2019 and 2020/2021, the counterfactual loss of the United States was almost a quarter smaller than its observed loss. Yet, the overall ranking of countries changed little. I conclude that looking at changes in mortality rates is essential, especially when populations with different levels of mortality are compared.

\medskip
\noindent{\sffamily\bfseries\color{accent}Keywords:} Life Expectancy Changes, Life Expectancy Differentials, Counterfactual Changes, Decomposition, Formal Demography
\end{abstract}

\section{Introduction}\label{sec:introduction}

Changes in life expectancy at birth are arguably the most widely used summary measure of population health to evaluate whether public health has improved or deteriorated. Statistical offices publish life expectancy figures routinely, and epidemiologists, health economists and demographers use the measure to compare populations and to assess the effects of health policies. During the COVID-19 years, for example, losses in life expectancy at birth were compared across countries and regions, and differences were often attributed to differences in public health measures \citep{Islam2021, Schoeley2022, Bonnet2024}. Similarly, changes in life expectancy at birth are compared between women and men, socioeconomic groups or regions, and the differences are interpreted in terms of health behaviour, access to health care, or policy \citep{Leon2011, Mackenbach2016, Sauerberg2024}. In addition, the development of life expectancy at birth is often used in the discussion of an appropriate retirement age, where slowing gains in life expectancy are easily read as slowing improvements in mortality. Consider a hypothetical scenario in which death rates at ages 50 to 80 decline by 5\% every year, starting from the mortality level of Germany in 2019, while mortality at all other ages remains unchanged. Although the relative improvement in mortality is the same in every year, life expectancy at birth would increase by 0.28 years (3.4 months) in the first year, but only by 0.20 years in the tenth year, 0.13 years in the twentieth year and 0.08 years (1.0 month) in the thirtieth year. Over the first decade, life expectancy at birth would increase by 2.4 years, over the third decade by only 1.0 year.\footnote{Both sexes combined; own calculations based on data from the \citet{HMD2026}. After 30 years, the death rates at ages 50 to 80 would be reduced to 21\% of their initial level, and life expectancy at birth would increase from 81.1 to 86.2 years.} Looking at the gains in life expectancy alone, mortality improvements would seem to slow down considerably, although death rates decline at a constant pace. The reason for that is, as mortality at these ages falls, fewer deaths remain to be saved, leading to the situation that the same relative decline in mortality yields smaller gains in life expectancy. 

Health policies and socioeconomic factors certainly contribute to differences in the development of life expectancy at birth. Nevertheless, a change in life expectancy at birth does not only reflect a change in mortality rates. The magnitude depends on the mortality schedule from which a population starts. Thus, gains or losses of populations with very different mortality levels can differ even though mortality rates developed similarly. For instance, life expectancy at birth in Lithuania increased by 3.7 years between 2009 and 2019 (from 72.9 to 76.5 years). Had Sweden experienced exactly the same relative changes in age-specific mortality rates, its life expectancy at birth would have increased by only 2.1 years (Sweden's life expectancy at birth value was 81.4 years in the year 2009). This is a 76\% larger gain for Lithuania as compared to Sweden, simply because mortality in Lithuania started from a much lower life expectancy level.\footnote{Both sexes combined. The figure for Sweden is obtained by applying the observed relative changes in age-specific death rates in Lithuania between 2009 and 2019 to Swedish death rates in 2009 \citep[data:][]{HMD2026}. The observed gain in Sweden was 1.7 years.} It is even possible to construct a scenario where the differences in death rates between two populations have decreased but the life expectancy at birth differential has widened \citep{Pollard1982}. 

The non-linear relationship between changes in mortality and changes in life expectancy is well known in mathematical demography \citep{Keyfitz1977, Pollard1982, Vaupel1986, Vaupel2003, Aburto2020}. For the sex gap in life expectancy, \citet{Glei2007} showed that this non-linearity matters empirically. They decomposed changes in the gap into an effect of changing sex ratios in mortality and an effect of sex differences in the age pattern of mortality, and found that the narrowing of the gap in most high-income countries since the 1970s resulted largely from the latter, i.e., the same rate of mortality decline produces smaller gains for women because their deaths are less dispersed across age. \citet{Cui2019} elaborated on this decomposition and examined the change in the sex gap in life expectancy at birth as well. Their added results emphasize the importance of considering differences in the initial mortality schedule rather than looking exclusively at differences in the progress of reducing age-specific death rates when explaining the narrowing of the sex gap in life expectancy. Despite these contributions, changes in life expectancy are still routinely compared without regard to differences in the initial mortality schedule, and no counterpart to the age-standardized death rate is commonly used for changes in life expectancy.

Because most of the previous literature is rather technical, its implications for applied work have received little attention. Put provocatively, comparing gains or losses in life expectancy at birth between populations with different initial mortality is, from a statistical point of view, comparable to comparing crude death rates between populations with different age structures. The comparison is not exact. Unlike the crude death rate, life expectancy at birth does not depend on the age structure of an observed population, because its weights are determined by mortality itself. Yet, the comparison highlights a mathematical similarity, i.e., in both cases, age-specific mortality, or its change, is summed with population-specific weights.

This paper has three aims. First, I describe formally the similarity between changes in life expectancy at birth and crude death rates. Second, similar to \citet{TimoninCooley2024}, who applied the relative changes in age-specific death rates of the United States to the mortality schedules of other countries, I compute counterfactual changes in life expectancy at birth, i.e., the changes that a common reference population would have experienced under the observed relative changes in mortality of each population. These counterfactual changes are the analogue of age-standardized death rates. Third, I propose a decomposition that splits the difference between two changes in life expectancy at birth into a mortality effect and a sensitivity effect. In the empirical part of this paper, I use data from the Human Mortality Database to examine whether counterfactual changes in life expectancy at birth lead to different conclusions in three popular research topics in epidemiology: sex differences in mortality, the convergence of mortality between East and West Germany, and losses in life expectancy at birth during the COVID-19 years.

\section{Methods}\label{sec:methods}

\subsection{Changes in life expectancy at birth}\label{subsec:changeLE}

Let $\mu(x,t)$ denote the force of mortality at age $x$ and time $t$. The life table survivorship function and the remaining life expectancy at age $x$ are determined entirely by $\mu(x,t)$:
\begin{equation}
  \begin{aligned}
  \ell(x,t) &= \exp\left(-\int_0^x \mu(a,t)\dd a\right),\\
  e(x,t) &= \frac{1}{\ell(x,t)}\int_x^{\omega} \ell(a,t)\dd a
         =\int_x^{\omega} \exp\left(-\int_x^{a} \mu(u,t)\dd u\right)\dd a ,
  \end{aligned}
  \label{eq:lt}
\end{equation}
where $\omega$ is the last open age interval and $e_0(t)=e(0,t)$ is life expectancy at birth. As shown by \citet{Keyfitz1977} and \citet{Vaupel2003}, the change in $e_0$ over time can be written as
\begin{equation}
  \dot{e}_0(t) = \frac{\partial e_0(t)}{\partial t}
  = -\int_0^{\omega} \mu(x,t)\,\ell(x,t)\,e(x,t)\,\rho(x,t)\dd x,
  \qquad
  \rho(x,t) = \frac{\partial \ln \mu(x,t)}{\partial t},
  \label{eq:e0dot}
\end{equation}
where $\rho(x,t)$ is the relative rate of change in the force of mortality at age $x$. A positive $\rho(x,t)$ indicates rising and a negative $\rho(x,t)$ falling mortality.

Since $\ell(x,t)$ and $e(x,t)$ are themselves functions of $\mu(x,t)$, the change in $e_0$ depends only on mortality. Mortality, however, enters in two different roles,
\begin{equation}
  \dot{e}_0(t)
  = -\int_0^{\omega}
    \underbrace{w(x,t)}_{\substack{\text{depends only on}\\ \text{mortality at time } t}}
    \;
    \underbrace{\rho(x,t)}_{\substack{\text{change in}\\ \text{mortality}}}
    \dd x,
  \qquad
  w(x,t) = \mu(x,t)\,\ell(x,t)\,e(x,t) = d(x,t)\,e(x,t),
  \label{eq:e0dot_w}
\end{equation}
where $d(x,t)=\mu(x,t)\ell(x,t)$ is the life table distribution of deaths. The weight $w(x,t)$ depends only on the mortality schedule at time $t$ and not on its change. The expression reveals that the same changes in mortality, $\rho(x,t)$, result in different changes in $e_0$ whenever populations differ in their mortality schedule. Integrating the weighting function over age gives a measure of life disparity, which has been called $e$-dagger, $e^\dagger(t)=\int_0^\omega d(x,t)\, e(x,t)\dd x$ \citep{Vaupel2003}.

\subsection{Counterfactual changes in life expectancy at birth}\label{subsec:standardization}

Eq.~\eqref{eq:e0dot_w} has the same structure as the crude death rate, which weights age-specific mortality rates by the population age structure $c(x,t)$. Age standardization replaces $c(x,t)$ with a standard age structure $c^s(x)$:
\begin{equation}
  \mathrm{CDR}(t) = \int_0^{\omega}
    \underbrace{c(x,t)}_{\substack{\text{population}\\ \text{age structure}}}
    \underbrace{\mu(x,t)}_{\substack{\text{age-specific}\\ \text{mortality}}}
    \dd x,
  \qquad
  \mathrm{ASDR}(t) = \int_0^{\omega}
    \underbrace{c^s(x)}_{\substack{\text{standard}\\ \text{age structure}}}
    \underbrace{\mu(x,t)}_{\substack{\text{age-specific}\\ \text{mortality}}}
    \dd x .
  \label{eq:asdr}
\end{equation}
In the same way, the change in life expectancy at birth is a ``crude'' measure that combines the change in mortality with population-specific weights. It is useful to ask how life expectancy at birth would have changed in a reference population with mortality schedule $\mu^r(x)$ if its death rates had changed by the same proportions as those of the population under study. Replacing $w(x,t)$ by the weights $w^r(x)=\mu^r(x)\,\ell^r(x)\,e^r(x)$ of the reference population gives the counterfactual change in life expectancy at birth,
\begin{equation}
  \dot{e}_0(t) = -\int_0^{\omega}
    \underbrace{w(x,t)}_{\substack{\text{population}\\ \text{weights}}}
    \underbrace{\rho(x,t)}_{\substack{\text{change in}\\ \text{mortality}}}
    \dd x,
  \qquad
  \dot{e}_0^{\,\mathrm{cf}}(t) = -\int_0^{\omega}
    \underbrace{w^r(x)}_{\substack{\text{reference}\\ \text{weights}}}
    \underbrace{\rho(x,t)}_{\substack{\text{change in}\\ \text{mortality}}}
    \dd x .
  \label{eq:e0dot_std}
\end{equation}
For an observed period from $t_1$ to $t_2$, with $\rho(x)=\ln\mu(x,t_2)-\ln\mu(x,t_1)$, the counterfactual change is
\begin{equation}
  \Delta e_0^{\mathrm{cf}} = -\int_0^{\omega} w^r(x)\,\rho(x)\dd x .
  \label{eq:Delta_e0_std}
\end{equation}
$\Delta e_0^{\mathrm{cf}}$ is, to first order, the change in $e_0$ the reference population would experience if its mortality rates changed by the same proportions as those of the population under study. Because all populations are evaluated with the same weights, differences in $\Delta e_0^{\mathrm{cf}}$ reflect differences in the change in mortality only. The counterfactual change thus plays the same role for changes in life expectancy as the age-standardized death rate does for death rates.

As with age-standardized death rates, the choice of the reference depends on the comparison. For two populations, I use the average of their weights, for several populations the weights of their pooled life table, and for one population over time the weights of a fixed reference year. Appendix~\ref{app:standardization} discusses the choice of the reference population in more detail.

\subsection{Decomposing the changes between populations into mortality and sensitivity effects}\label{sec:comparison}
The following decomposition method is equivalent to the decomposition of changes in the sex gap in life expectancy proposed by \citet{Glei2007}. While they obtained it by expressing male and female death rates as functions of their ratio and their geometric mean, I present the decomposition in the form of the Kitagawa decomposition of crude death rates \citep{Kitagawa1955, Preston2001}. This allows making the analogy to the rate effect and the composition effect, which makes its interpretation intuitive for any pair of populations. In addition, it reveals that the mortality effect equals the difference between counterfactual changes in life expectancy, linking the decomposition to the counterfactual changes introduced in Section~\ref{subsec:standardization}.

For two populations $A$ and $B$, the difference between their annual changes at age $x$ can be split into
\begin{equation}
  -\bar{w}_{A,t}(x)\,\rho_{A,t}(x) + \bar{w}_{B,t}(x)\,\rho_{B,t}(x) = M_t(x) + S_t(x),
  \label{eq:kitagawa}
\end{equation}
with
\begin{align}
  M_t(x) &= -\underbrace{\tfrac{1}{2}\big(\bar{w}_{A,t}(x)+\bar{w}_{B,t}(x)\big)}_{\text{average weight}}
          \;\underbrace{\big(\rho_{A,t}(x)-\rho_{B,t}(x)\big)}_{\substack{\text{difference in}\\ \text{mortality change}}},
  \label{eq:M}\\[4pt]
  S_t(x) &= -\underbrace{\tfrac{1}{2}\big(\rho_{A,t}(x)+\rho_{B,t}(x)\big)}_{\text{average mortality change}}
          \;\underbrace{\big(\bar{w}_{A,t}(x)-\bar{w}_{B,t}(x)\big)}_{\text{difference in weights}} .
  \label{eq:S}
\end{align}
where $\bar{w}_t(x)$ is the weight $w(x,t)$ averaged over the year. I use $\bar{w}_t(x)$ because Eq.~\eqref{eq:e0dot_w} describes instantaneous changes, but empirical data refer to discrete years. I therefore compute changes from one year to the next and sum them over the period of interest. For the change from year $t$ to $t+1$, with $\rho_t(x)=\ln\mu(x,t+1)-\ln\mu(x,t)$,
\begin{equation}
  \Delta e_{0,t} = e_0(t+1) - e_0(t)
  = -\int_0^{\omega} \bar{w}_t(x)\,\rho_t(x)\dd x .
  \label{eq:Delta_e0}
\end{equation}
Under the assumption that mortality changes at a constant relative rate within each year, Eq.~\eqref{eq:Delta_e0} holds as shown by \citet{Horiuchi2008}, and $\bar{w}_t(x)$ is the weight averaged over the year (Appendix~\ref{app:observed}).

The \emph{mortality effect} $M_t(x)$ corresponds to the rate effect, i.e., the part of the difference due to different changes in mortality. The \emph{sensitivity effect} $S_t(x)$ corresponds to the composition effect, i.e., the part due to different weights. This effect can be further disentangled into a level effect and an age-pattern effect, as I show in Appendix~\ref{app:sensitivity}. Summing over ages and years gives
\begin{equation}
  \Delta e_0^A - \Delta e_0^B
  = \underbrace{\sum_{t=t_1}^{t_2-1}\int_0^{\omega} M_t(x)\dd x}_{\text{mortality effect}}
  + \underbrace{\sum_{t=t_1}^{t_2-1}\int_0^{\omega} S_t(x)\dd x}_{\text{sensitivity effect}} ,
  \label{eq:total_decomp}
\end{equation}
and both effects can equally be summed over any age group. Only the mortality effect allows making statements such as mortality developed more favourably in one population than in the other. If both populations experience the same change in mortality, the mortality effect is zero and the entire difference in $\Delta e_0$ is a sensitivity effect.

As the rate effect equals the difference between age-standardized death rates when the average age structure is used as the standard, the mortality effect equals the difference between counterfactual changes when the average weights are used as the reference,
\begin{equation}
  \int_0^{\omega} M_t(x)\dd x = \Delta e_{0,t}^{\mathrm{cf},A} - \Delta e_{0,t}^{\mathrm{cf},B},
  \qquad
  w^r(x) = \tfrac12\big(\bar{w}_{A,t}(x)+\bar{w}_{B,t}(x)\big).
  \label{eq:M_std}
\end{equation}
The sensitivity effect is the part of the difference that disappears when both populations are evaluated with the same weights.

\section{Results}\label{sec:results}

\subsection{Data and computation}\label{subsec:data}

All analyses are based on data from the \citet{HMD2026}. For each population and calendar year, I use death counts and exposures by single year of age, aggregate ages 100 and above into an open age interval, and compute death rates as $m_x=D_x/E_x$. Life tables are constructed using standard methods as described by \citet{Preston2001}. From these life tables, I compute three quantities. First, for each pair of consecutive years, the observed change in life expectancy at birth, $\Delta e_{0,t}$, together with its age contributions $-\bar{w}_{x,t}\rho_{x,t}$ (Eq.~\eqref{eq:Delta_e0}), where the averaged weights $\bar{w}_{x,t}$ are computed as described in Appendix~\ref{app:lifetables}. Changes over longer periods are the sums of the annual changes. Second, the counterfactual change in life expectancy at birth, $\Delta e_0^{\mathrm{cf}}=-\sum_x w^r_x\rho_x$ (Eq.~\eqref{eq:Delta_e0_std}), where $\rho_x$ is the log change in death rates between the first and the last year of the period. In line with Section~\ref{subsec:standardization}, the reference weights are the average weights of women and men for the sex comparison, the average weights of East and West Germany for the comparison of the two parts of Germany, and the weights of the pooled life table of all 27 countries in 2019 for the COVID-19 analysis, which refers to the changes from 2019 to 2020 and from 2019 to 2021. Third, the decomposition of differences in $\Delta e_0$ between two populations into the mortality and the sensitivity effect (Eq.~\eqref{eq:total_decomp}), which I present for different age groups.

For the sex comparison, I use eight countries with large populations and complete data for 1990--2019 (the HMD series for Germany as a whole starts in 1990): Germany, Spain, France, England and Wales, Italy, Japan, Poland and the United States. For the comparison of East and West Germany, I use the HMD series for the two parts of the country from 1990 to 2019. For the COVID-19 analysis, I start from 31 countries with data for 2019--2021 and exclude those in which at least one single-year age group recorded no deaths in one of these years, because $\rho_x$ is then undefined. This excludes Estonia, Finland, Ireland and Latvia and leaves 27 countries.

\subsection{The weights}\label{subsec:res_weights}

\begin{figure}[!htbp]
  \centering
  \includegraphics[width=\textwidth]{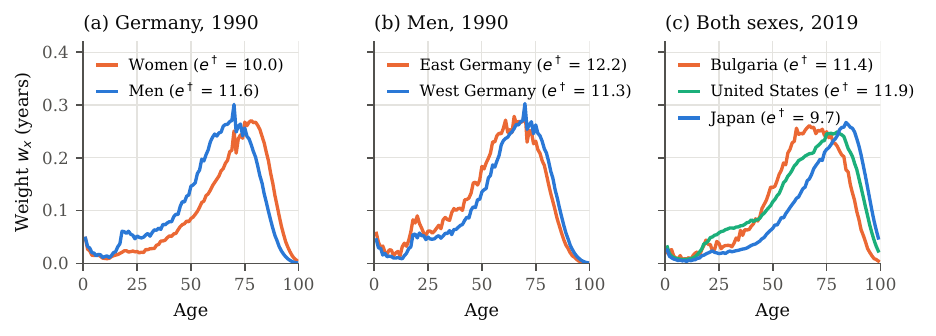}
  \caption{Weights $w_x$ by age for selected populations. The weight at age $x$ is the change in $e_0$, in years, caused by a one-unit change in the logarithm of the death rate at age $x$. $e^\dagger$ is life disparity, the sum of the weights over age. Age~0 and the open age interval 100+ are not shown.\\ \emph{Source:} Human Mortality Database, own calculations.}
  \label{fig:weights}
\end{figure}

Figure~\ref{fig:weights} shows the weights $w_x$ for the populations compared in the three applications. The weights are small in childhood, rise with age and peak in the late 60s to the early 80s, where most life table deaths occur, before they fall to zero at the highest ages. The populations differ in both the level and the age pattern of their weights. In 1990, German men had larger weights than German women at all ages below about 75 years, especially at young adult and middle ages, whereas women had larger weights at older ages (panel a). Their life disparity was 11.6 years compared with 10.0 years for women. Hence, a uniform decline in mortality of one percent at all ages would have raised the life expectancy of men by about 0.116 years and that of women by about 0.100 years, i.e., by 15\% more.\footnote{Following \citet{Keyfitz1977}, a small relative change $\rho$ that is the same at all ages changes life expectancy at birth by $\Delta e_0 \approx -\rho\, e^\dagger$ (Eq.~\eqref{eq:e0dot_w} with $\rho(x,t)=\rho$ for all $x$). With $\rho=-0.01$, the gain is one percent of $e^\dagger$, i.e., $0.01\times 11.6\approx0.116$ years for men and $0.01\times10.0\approx0.100$ years for women.} East German men had a higher life disparity than West German men in 1990 (12.2 vs.\ 11.3 years) and larger weights at most ages below 70 (panel b). Among the countries in the COVID-19 analysis (panel c), the weights of the United States and Bulgaria are considerably larger than those of Japan between ages 20 and 75, whereas Japan has the largest weights above age 85. A uniform ten percent increase in mortality would, to first order, reduce life expectancy at birth by 1.19 years in the United States, by 1.14 years in Bulgaria, but by only 0.97 years in Japan.

\subsection{Sex differences in life expectancy gains}\label{subsec:res_sex}

Table~\ref{tab:sex} compares the changes in life expectancy at birth of women and men between 1990 and 2019. In seven of the eight countries, men gained more life expectancy than women, by between 1.35 (Poland) and 2.40 years (France), so that the sex gap in $e_0$ narrowed. Japan is the only exception, with slightly larger gains for women. Such results are usually read as evidence that mortality improved faster among men than among women.

The decomposition shows that this interpretation holds only partially. In all countries, the sensitivity effect is negative, i.e., it favours men. Because men's weights are larger, the same relative decline in mortality yields larger gains in life expectancy for men than for women. The size of the sensitivity effect ranges from $-0.38$ years in the United States to $-1.72$ years in Spain. It is mainly a level effect, reflecting the higher life disparity of men, while the part due to the different age pattern of the weights is smaller. In Spain and Poland, the entire difference between women and men is a sensitivity effect. The mortality effect is close to zero or even slightly positive (0.05 and 0.24 years), which means that women's mortality declined at least as fast as men's. Japan shows the opposite case. Women's mortality declined considerably faster than men's (mortality effect 1.00 years), but this was almost fully offset by the larger sensitivity of men ($-0.85$ years), so that the observed gains were nearly equal. Only in the United States and in England and Wales does the mortality effect account for most of the larger gains of men.

The counterfactual differences (last three columns of Table~\ref{tab:sex}) lead to the same overall conclusions. With the average weights of women and men in 1990 as the reference, the advantage of men shrinks from 1.73 to 0.82 years in Germany and from 1.68 to 0.33 years in Spain, vanishes in Poland and turns into an advantage of women of 0.84 years in Japan. Over almost three decades, the weights shift to higher ages as mortality declines, so that the two measures weight the age-specific changes slightly differently. The choice of the reference also matters. With the weights of men as the reference, the advantage of men in the counterfactual changes is larger than with the weights of women, in all countries. In Spain and Poland, the sign of the counterfactual difference depends on the reference (Spain: $0.12$ years with women's and $-0.77$ years with men's weights; Poland: $0.13$ and $-0.13$ years), which again indicates that the changes in mortality of women and men were similar there. In the other six countries, the direction of the counterfactual difference does not depend on the reference.

\begin{table}[!htbp]
  \centering
  \caption{Changes in life expectancy at birth of women and men, 1990--2019, and decomposition of their difference (women minus men) into the mortality effect $M$ and the sensitivity effect $S$, in years.}
  \label{tab:sex}
  \footnotesize
  \setlength{\tabcolsep}{3.5pt}
  \begin{tabular}{l rrr rrrr rrr}
\toprule
 & \multicolumn{3}{c}{Observed change in $e_0$} & \multicolumn{4}{c}{Decomposition of difference} & \multicolumn{3}{c}{Counterfactual difference} \\
\cmidrule(lr){2-4}\cmidrule(lr){5-8}\cmidrule(lr){9-11}
Country & Women & Men & Diff. & $M$ & $S$ & $S$ level & $S$ pattern & Average & Women & Men \\
\midrule
Germany & 5.08 & 6.82 & $-$1.73 & $-$0.63 & $-$1.10 & $-$0.84 & $-$0.26 & $-$0.82 & $-$0.62 & $-$1.03 \\
Spain & 5.62 & 7.29 & $-$1.68 & 0.05 & $-$1.72 & $-$1.26 & $-$0.47 & $-$0.33 & 0.12 & $-$0.77 \\
France & 4.64 & 7.04 & $-$2.40 & $-$1.12 & $-$1.28 & $-$1.07 & $-$0.21 & $-$1.29 & $-$0.87 & $-$1.71 \\
England \& Wales & 4.74 & 6.66 & $-$1.92 & $-$1.28 & $-$0.65 & $-$0.42 & $-$0.22 & $-$1.36 & $-$1.30 & $-$1.41 \\
Italy & 5.16 & 7.51 & $-$2.35 & $-$0.89 & $-$1.46 & $-$0.86 & $-$0.61 & $-$1.16 & $-$0.80 & $-$1.52 \\
Japan & 5.58 & 5.42 & 0.16 & 1.00 & $-$0.85 & $-$0.72 & $-$0.13 & 0.84 & 1.02 & 0.65 \\
Poland & 6.38 & 7.73 & $-$1.35 & 0.24 & $-$1.59 & $-$1.47 & $-$0.12 & 0.00 & 0.13 & $-$0.13 \\
United States & 2.76 & 4.67 & $-$1.91 & $-$1.53 & $-$0.38 & $-$0.42 & 0.04 & $-$1.51 & $-$1.41 & $-$1.62 \\
\bottomrule
\end{tabular}
  \par\medskip
  \begin{minipage}{\textwidth}\scriptsize
    \emph{Notes:} Diff.\ $=\Delta e_0^{\text{Women}}-\Delta e_0^{\text{Men}}=M+S$ (Eq.~\eqref{eq:total_decomp}). $S$ level and $S$ pattern: parts of $S$ due to differences in life disparity and in the age pattern of the weights (Eq.~\eqref{eq:S_split}). Counterfactual difference: $\Delta e_0^{\mathrm{cf},\text{Women}}-\Delta e_0^{\mathrm{cf},\text{Men}}$ with the average weights of women and men in 1990, the weights of women, or the weights of men in 1990 as the reference. \emph{Source:} Human Mortality Database, own calculations.
  \end{minipage}
\end{table}

\begin{figure}[!htbp]
  \centering
  \includegraphics[width=\textwidth]{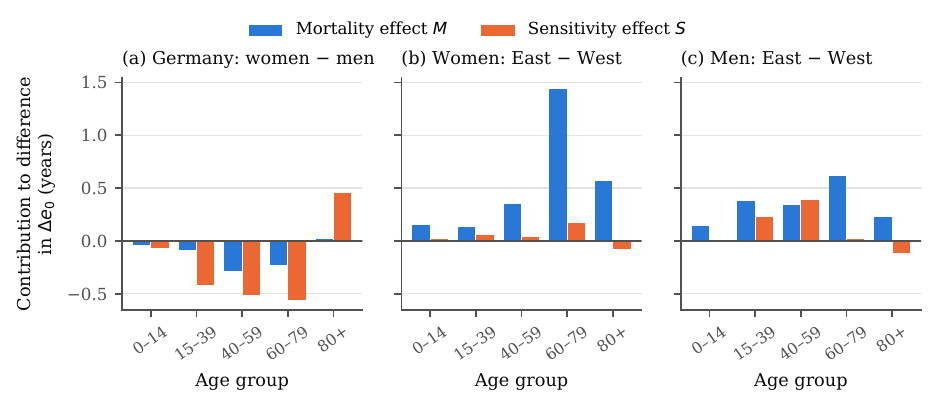}
  \caption{Decomposition of differences in the change in life expectancy at birth, 1990--2019, into the mortality effect $M$ and the sensitivity effect $S$, by age group. (a) Women minus men in Germany. (b) and (c) East minus West Germany for women and men. The contributions of all age groups sum to the totals in Tables~\ref{tab:sex} and~\ref{tab:eastwest}.\\ \emph{Source:} Human Mortality Database, own calculations.}
  \label{fig:agegroups}
\end{figure}

Panel (a) of Figure~\ref{fig:agegroups} shows the decomposition by age group for Germany. The sensitivity effect is negative below age 80, where men have larger weights, in particular at ages 15--79, and positive at ages 80 and above, where women have larger weights. The mortality effect is negative mainly at ages 40--79, where men's mortality declined faster than women's. Overall, about two thirds of the 1.73-year larger gain of German men ($-1.10$ years) are due to the sensitivity effect and one third ($-0.63$ years) to faster declines in men's mortality.

\subsection{The convergence of East and West Germany}\label{subsec:res_eastwest}

\begin{figure}[!htbp]
  \centering
  \includegraphics[width=\textwidth]{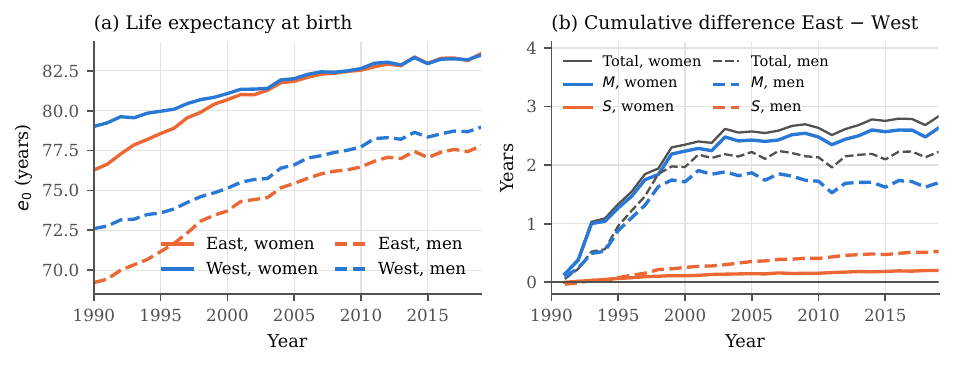}
  \caption{East and West Germany, 1990--2019. (a) Life expectancy at birth by sex. (b) Cumulative difference in the change in life expectancy at birth since 1990 (East minus West), and its decomposition into the mortality effect $M$ and the sensitivity effect $S$. Solid lines: women; dashed lines: men.\\ \emph{Source:} Human Mortality Database, own calculations.}
  \label{fig:eastwest}
\end{figure}

In 1990, life expectancy at birth in East Germany was 2.7 years lower than in West Germany for women (76.3 vs.\ 79.0 years) and 3.4 years lower for men (69.2 vs.\ 72.6 years). By 2019, East German women had caught up completely (83.6 vs.\ 83.5 years), whereas a gap of 1.2 years remained for men (77.8 vs.\ 79.0 years; Figure~\ref{fig:eastwest}a). East German women gained 2.84 years more than West German women and East German men 2.23 years more than West German men (Table~\ref{tab:eastwest}).

Because the weights of East Germans were larger, part of this convergence would have occurred even if mortality had declined at the same pace in both parts of the country. For women, this sensitivity effect is small (0.20 years), and 93\% of the larger gain in the East is due to the faster decline in mortality. For men, the sensitivity effect amounts to 0.53 years, i.e., almost a quarter of the larger gain of East German men reflects their higher initial mortality and the resulting higher sensitivity of their life expectancy. As Figure~\ref{fig:agegroups} (panels b and c) shows, the sensitivity effect for men arises at ages 15--59, where the weights of East German men were particularly large after reunification. The mortality effect for women is concentrated at ages 60 and above, with 1.44 years at ages 60--79 alone, whereas for men it is spread more evenly over the ages 15--79.

The development over time adds a further insight (Figure~\ref{fig:eastwest}b). The mortality effect accumulates almost entirely during the 1990s. For men, it reaches its maximum of 1.91 years in 2001 and declines afterwards to 1.70 years in 2019. In terms of the change in mortality, East German men have thus fallen slightly behind West German men since the early 2000s. The sensitivity effect, by contrast, continues to grow as long as mortality declines in both populations and rises from 0.27 years in 2001 to 0.53 years in 2019. As a result, the observed difference in gains has remained roughly constant for men since 2001. This stability, however, conceals that mortality has declined slightly faster in West Germany, and that East German men kept pace only because the same relative improvements translate into larger gains at their higher level of mortality. For women, the mortality effect increases only slightly after 2003 (from 2.48 to 2.63 years), and the sensitivity effect remains small.

The counterfactual differences confirm these results. With the average weights of 1990 as the reference, East Germany gained 2.93 years more than West Germany for women and 1.86 years more for men. With the weights of either population as the reference, the counterfactual differences range from 2.80 to 3.06 years for women and from 1.69 to 2.02 years for men, so that the conclusions do not depend on the choice of the reference.

\begin{table}[!htbp]
  \centering
  \caption{Changes in life expectancy at birth in East and West Germany, 1990--2019, and decomposition of their difference (East minus West) into the mortality effect $M$ and the sensitivity effect $S$, in years.}
  \label{tab:eastwest}
  \footnotesize
  \setlength{\tabcolsep}{3.5pt}
  \begin{tabular}{l rrr rrrr rrr}
\toprule
 & \multicolumn{3}{c}{Observed change in $e_0$} & \multicolumn{4}{c}{Decomposition of difference} & \multicolumn{3}{c}{Counterfactual difference} \\
\cmidrule(lr){2-4}\cmidrule(lr){5-8}\cmidrule(lr){9-11}
Sex & East & West & Diff. & $M$ & $S$ & $S$ level & $S$ pattern & Average & East & West \\
\midrule
Women & 7.31 & 4.47 & 2.84 & 2.63 & 0.20 & 0.08 & 0.12 & 2.93 & 3.06 & 2.80 \\
Men & 8.59 & 6.36 & 2.23 & 1.70 & 0.53 & 0.41 & 0.11 & 1.86 & 2.02 & 1.69 \\
Total & 7.76 & 5.24 & 2.52 & 2.13 & 0.38 & 0.28 & 0.11 & 2.29 & 2.40 & 2.19 \\
\bottomrule
\end{tabular}
  \par\medskip
  \begin{minipage}{\textwidth}\scriptsize
    \emph{Notes:} Diff.\ $=\Delta e_0^{\text{East}}-\Delta e_0^{\text{West}}=M+S$. Counterfactual difference: $\Delta e_0^{\mathrm{cf},\text{East}}-\Delta e_0^{\mathrm{cf},\text{West}}$ with the average weights of East and West Germany in 1990, the weights of East Germany, or the weights of West Germany in 1990 as the reference. See also the notes to Table~\ref{tab:sex}. \emph{Source:} Human Mortality Database, own calculations.
  \end{minipage}
\end{table}

\subsection{Losses in life expectancy between 2019 and 2020 or 2021}\label{subsec:res_covid}

\begin{figure}[!htbp]
  \centering
  \includegraphics[width=\textwidth]{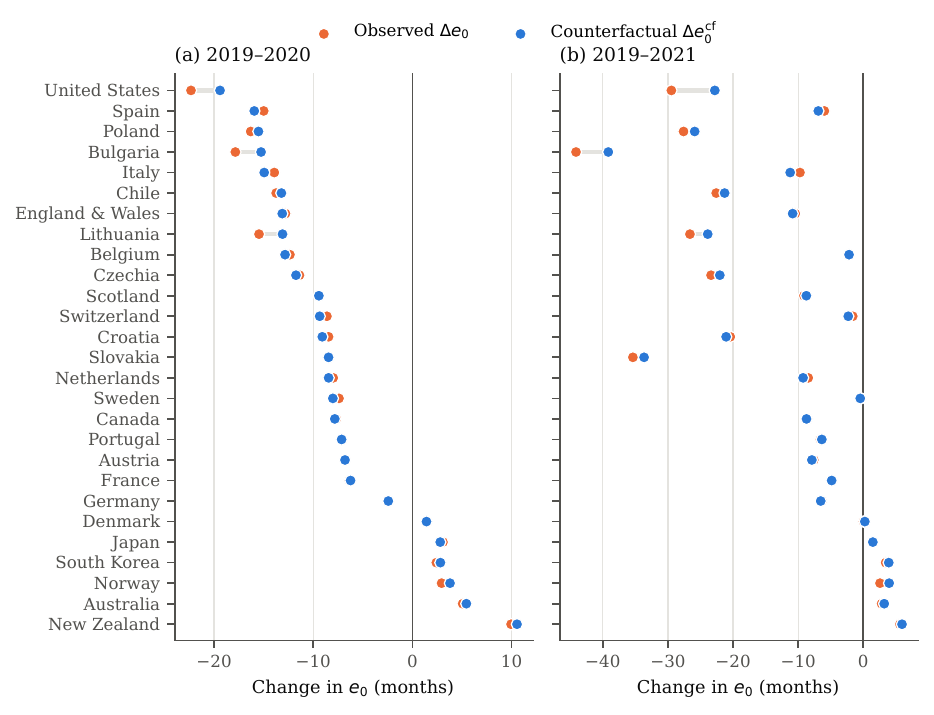}
  \caption{Observed and counterfactual changes in life expectancy at birth, both sexes, in months, (a) 2019--2020 and (b) 2019--2021. Reference: weights of the pooled life table of all 27 countries in 2019. Countries are sorted by the counterfactual change 2019--2020 in both panels.\\ \emph{Source:} Human Mortality Database, own calculations.}
  \label{fig:covid}
\end{figure}

Table~\ref{tab:covid} and Figure~\ref{fig:covid} show the observed and the counterfactual changes in life expectancy at birth from 2019 to 2020 and from 2019 to 2021 for the 27 countries. I compare the years 2020 and 2021 with the pre-COVID-19 level of 2019, because the change from 2020 to 2021 partly reflects a recovery from the losses of 2020 and is therefore hard to interpret on its own. Because the changes are small compared with the level of life expectancy, I report them in months rather than years. The pooled life table of 2019, which serves as the reference, has a life expectancy at birth of 81.3 years and a life disparity of 10.7 years. From 2019 to 2020, life expectancy at birth declined in 21 of the 27 countries, most strongly in the United States ($-22.3$ months), Bulgaria ($-17.9$ months), Poland ($-16.3$ months), Lithuania ($-15.5$ months) and Spain ($-15.0$ months). By 2021, the losses had grown further in Central and Eastern Europe and in the United States, whereas most Western European countries had partly recovered. Compared with 2019, life expectancy at birth in 2021 was lower in 21 countries, by up to 44.1 months in Bulgaria and 35.4 months in Slovakia.

\begin{table}[!htbp]
  \centering
  \caption{Observed ($\Delta e_0$) and counterfactual ($\Delta e_0^{\mathrm{cf}}$) changes in life expectancy at birth from 2019 to 2020 and from 2019 to 2021, both sexes, in months, with ranks (1 = largest loss).}
  \label{tab:covid}
  \footnotesize
  \setlength{\tabcolsep}{4pt}
  \begin{tabular}{l rr rrrr rrrr}
\toprule
 & & & \multicolumn{4}{c}{2019--2020} & \multicolumn{4}{c}{2019--2021} \\
\cmidrule(lr){4-7}\cmidrule(lr){8-11}
Country & $e_0$ & $e^\dagger$ & $\Delta e_0$ & $\Delta e_0^{\mathrm{cf}}$ & Rank & Rank$^{\mathrm{cf}}$ & $\Delta e_0$ & $\Delta e_0^{\mathrm{cf}}$ & Rank & Rank$^{\mathrm{cf}}$ \\
\midrule
United States & 79.1 & 11.9 & $-$22.3 & $-$19.4 & 1 & 1 & $-$29.5 & $-$22.8 & 3 & 5 \\
Spain & 83.5 & 9.6 & $-$15.0 & $-$15.9 & 5 & 2 & $-$6.0 & $-$6.9 & 17 & 15 \\
Poland & 77.8 & 11.4 & $-$16.3 & $-$15.5 & 3 & 3 & $-$27.6 & $-$25.9 & 4 & 3 \\
Bulgaria & 75.1 & 11.4 & $-$17.9 & $-$15.3 & 2 & 4 & $-$44.1 & $-$39.2 & 1 & 1 \\
Italy & 83.4 & 9.2 & $-$13.9 & $-$14.9 & 6 & 5 & $-$9.7 & $-$11.2 & 10 & 9 \\
Chile & 80.6 & 11.4 & $-$13.7 & $-$13.2 & 7 & 6 & $-$22.6 & $-$21.3 & 7 & 7 \\
England \& Wales & 81.6 & 10.1 & $-$12.8 & $-$13.1 & 8 & 7 & $-$10.5 & $-$10.9 & 9 & 10 \\
Lithuania & 76.5 & 11.9 & $-$15.5 & $-$13.1 & 4 & 8 & $-$26.6 & $-$23.9 & 5 & 4 \\
Belgium & 81.8 & 9.9 & $-$12.4 & $-$12.8 & 9 & 9 & $-$2.2 & $-$2.2 & 19 & 20 \\
Czechia & 79.2 & 10.2 & $-$11.4 & $-$11.7 & 10 & 10 & $-$23.4 & $-$22.0 & 6 & 6 \\
Scotland & 79.2 & 10.9 & $-$9.4 & $-$9.4 & 11 & 11 & $-$9.1 & $-$8.7 & 11 & 12 \\
Switzerland & 83.8 & 9.1 & $-$8.6 & $-$9.4 & 12 & 12 & $-$1.6 & $-$2.3 & 20 & 19 \\
Croatia & 78.3 & 10.2 & $-$8.5 & $-$9.1 & 13 & 13 & $-$20.4 & $-$21.1 & 8 & 8 \\
Slovakia & 77.5 & 10.9 & $-$8.4 & $-$8.5 & 14 & 14 & $-$35.4 & $-$33.7 & 2 & 2 \\
Netherlands & 82.1 & 9.4 & $-$8.0 & $-$8.4 & 15 & 15 & $-$8.5 & $-$9.2 & 13 & 11 \\
Sweden & 83.1 & 9.2 & $-$7.4 & $-$8.0 & 17 & 16 & $-$0.4 & $-$0.5 & 21 & 21 \\
Canada & 82.3 & 10.5 & $-$7.7 & $-$7.8 & 16 & 17 & $-$8.6 & $-$8.7 & 12 & 13 \\
Portugal & 82.0 & 10.1 & $-$7.1 & $-$7.1 & 18 & 18 & $-$6.6 & $-$6.4 & 15 & 17 \\
Austria & 81.9 & 9.7 & $-$6.8 & $-$6.8 & 19 & 19 & $-$7.6 & $-$7.9 & 14 & 14 \\
France & 82.8 & 10.4 & $-$6.3 & $-$6.2 & 20 & 20 & $-$4.8 & $-$4.9 & 18 & 18 \\
Germany & 81.1 & 9.9 & $-$2.4 & $-$2.4 & 21 & 21 & $-$6.4 & $-$6.5 & 16 & 16 \\
Denmark & 81.4 & 9.8 & 1.5 & 1.4 & 22 & 22 & 0.0 & 0.2 & 22 & 22 \\
Japan & 84.5 & 9.7 & 3.1 & 2.8 & 25 & 23 & 1.5 & 1.5 & 23 & 23 \\
South Korea & 83.5 & 9.7 & 2.4 & 2.8 & 23 & 24 & 3.4 & 3.9 & 26 & 25 \\
Norway & 83.0 & 9.3 & 2.9 & 3.8 & 24 & 25 & 2.6 & 4.0 & 24 & 26 \\
Australia & 83.2 & 9.7 & 5.1 & 5.4 & 26 & 26 & 2.8 & 3.2 & 25 & 24 \\
New Zealand & 81.8 & 10.0 & 10.0 & 10.5 & 27 & 27 & 5.6 & 6.0 & 27 & 27 \\
\midrule
Pooled reference & 81.3 & 10.7 & & & & & & & & \\
\bottomrule
\end{tabular}
  \par\medskip
  \begin{minipage}{\textwidth}\scriptsize
    \emph{Notes:} $e_0$ and $e^\dagger$: life expectancy at birth and life disparity in 2019. Reference: weights of the pooled life table of all 27 countries in 2019. Changes in months are changes in years multiplied by 12. Rank and Rank$^{\mathrm{cf}}$: rank of the observed and of the counterfactual change. Countries are sorted by the counterfactual change 2019--2020. \emph{Source:} Human Mortality Database, own calculations.
  \end{minipage}
\end{table}

For most Western European countries, whose life disparity is close to that of the reference, observed and counterfactual changes are almost identical. The differences are largest for countries with high life disparity. In the United States, the loss from 2019 to 2020 shrinks from 22.3 to 19.4 months in the counterfactual, in Bulgaria from 17.9 to 15.3 months and in Lithuania from 15.5 to 13.1 months. In countries with low life disparity, such as Spain, Italy and Switzerland, the counterfactual losses are larger than the observed ones. The difference between the observed and the counterfactual change is strongly correlated with life disparity in 2019 (Pearson's $r=-0.75$ for 2019--2020 and $-0.76$ for 2019--2021). This changes the ranking among the countries with the largest losses in 2020. Spain moves from the fifth to the second-largest loss, ahead of Bulgaria, which falls from second to fourth place, and Lithuania, which falls from fourth to eighth place. In terms of the change in mortality, the first year with COVID-19 was thus worse in Spain than in Bulgaria and Lithuania, although the observed losses in life expectancy suggest the opposite.

Over the two years from 2019 to 2021, the difference between observed and counterfactual changes is largest for the United States. Its observed loss of 29.5 months is reduced to 22.8 months, i.e., almost a quarter of the observed loss (6.7 months) reflects the high sensitivity of life expectancy in the United States, which arises from its comparatively high mortality at young and middle ages. In the counterfactual ranking, the United States falls from the third to the fifth-largest loss, behind Poland and Lithuania. Losses also shrink in Bulgaria (from 44.1 to 39.2 months) and Lithuania (from 26.6 to 23.9 months), whereas they increase in Italy (from 9.7 to 11.2 months) and Spain (from 6.0 to 6.9 months). Norway's gain rises from 2.6 to 4.0 months.

Taken as a whole, however, the ranking of countries changes little. The Spearman rank correlation between observed and counterfactual changes is 0.99 for both periods. The differences between countries in the increase in mortality during the 2020-2021 period were large compared with the differences in sensitivity, so that the counterfactual comparison changes the conclusions mainly where countries with different levels of mortality experienced similar losses. The results do not depend on the choice of the reference. When the weights of each of the 27 countries in 2019 are used as the reference instead of those of the pooled life table, the Spearman rank correlation with the counterfactual changes based on the pooled reference is at least 0.98 for both periods.

\section{Discussion}\label{sec:discussion}

Trends in life expectancy at birth and differences in these trends between populations are usually attributed to changes in mortality only. In this paper, I have shown that the same change in mortality can result in different changes in life expectancy at birth, depending on the mortality schedule from which a population starts. As noted in the Introduction, this is not a new finding. The dependence of changes in life expectancy on the initial level and age pattern of mortality has been discussed for the narrowing sex gap in life expectancy \citep{Glei2007, Cui2019} and for trends in life expectancy and life span equality more generally \citep{Aburto2020}. In the context of COVID-19, \citet{Goldstein2020} used Keyfitz's life table entropy to translate proportional increases in mortality into losses of life expectancy, which implies that the same increase in mortality costs more life expectancy in populations with higher entropy. The link between the life expectancy measure and age-standardization has been discussed previously by \citet{Modig2020} and \citet{Sauerberg2024MAD}. While \citet{Modig2020} showed that comparing levels of life expectancy implicitly compares crude death rates of different stationary populations, they did not propose a corrective measure. I extend their argument to changes in life expectancy by providing counterfactual changes in life expectancy, the analogue of age-standardized death rates, and a decomposition similar to the one used for crude death rate comparisons. The counterfactual changes build on \citet{TimoninCooley2024}, who applied the relative changes in age-specific death rates of the United States between 2010 and 2019 to the mortality schedules of 41 other countries and found markedly different changes in life expectancy at birth. 

The change in life expectancy at birth weights the age-specific changes in mortality by population-specific weights, just as the crude death rate weights age-specific mortality rates by the population age structure. Comparing observed changes in life expectancy between populations with different mortality levels is therefore analogous to comparing crude death rates between populations with different age structures. The analogy has its limits, however. In a classic age standardization, the age structure of a population is shaped not only by mortality but also by past fertility and migration, and differences in age structure between populations can be much larger than the differences in the weights I observe. In my applications, for example, the life disparity of German men exceeded that of German women by 15\% in 1990, and the life disparity of the 27 countries in the analysis of the years 2020 and 2021 ranged only from 9.1 years (Switzerland) to 11.9 years (Lithuania and the United States) in 2019. Evaluating changes in mortality with common weights therefore tends to matter less for changes in life expectancy than age standardization does for crude death rates.

The decomposition of differences in changes in life expectancy into a mortality effect and a sensitivity effect is the analogue of Kitagawa's decomposition of differences between crude death rates into a rate effect and a composition effect \citep{Kitagawa1955}. Presenting observed and counterfactual changes in life expectancy is meant to make the role of the initial mortality schedule comprehensible. It should not be understood as a claim that conventional changes in life expectancy at birth are distorted or wrong. Life expectancy at birth is a non-linear function of age-specific mortality, and its changes correctly describe how the average length of life in a population develops. My point is a different one. Changes in life expectancy at birth cannot be attributed entirely to changes in mortality, and a larger or smaller change in one population compared with another can therefore not be ascribed entirely to a more or less favourable development of mortality. This holds in particular when small differences in changes over short periods are compared between populations with different levels of mortality. Statistical offices routinely publish changes in life expectancy at birth between two consecutive years, and small differences in these changes between countries, regions or the sexes are often interpreted as the result of health policies or health behaviour, although they may partly reflect differences in the weights. Counterfactual changes isolate the change in mortality while keeping life years as the unit, which health policy makers and the media usually find more intuitive than differences in age-standardized death rates.

I demonstrated the relevance of the weights in three empirical examples. In line with \citet{Glei2007} and \citet{Cui2019}, I find that the larger gains in life expectancy of men compared with women between 1990 and 2019 are to a considerable extent a result of the different weights. In Germany, about two thirds of the larger gain of men are due to the sensitivity effect, and in Spain and Poland, the entire difference is a sensitivity effect. This finding, however, does not mean that there are no differences in the development of age- and cause-specific death rates between women and men (see for example, \citet{Sauerberg2023} for an analysis of age- and cause-specific mortality across regions of seven European countries). For the convergence of East and West Germany and for the losses in life expectancy during the COVID-19 years, the picture is more mixed. I find cases in which the counterfactual comparison matters. Almost a quarter of the larger gain of East German men compared with West German men reflects their higher initial mortality, and the stable difference in gains since the early 2000s conceals that mortality declined slightly faster in West Germany. Moreover, the counterfactual loss of the United States from 2019 to 2021 was almost a quarter smaller than its observed loss, and Spain moved from the fifth to the second-largest loss in 2020. Overall, however, the conclusions did not change much. The convergence of East and West Germany was mainly driven by faster mortality declines in the East, and the ranking of countries by their losses during the 2020/2021 period was almost identical for observed and counterfactual changes.

The relevance of counterfactual changes should not be limited to formal demography. Applied studies routinely compare changes in life expectancy between populations that start from very different levels of mortality. \citet{Woolf2021}, for example, looked at the development of life expectancy at birth in the United States and peer countries, and \citet{Andrasfay2021} projected that the losses in 2020 of the Black and the Latino population in the United States would be about three and four and a half times those of the White population, widening the Black--White gap in life expectancy substantially. In the comparison of the United States with its peer countries and of the Black with the White population, the populations with the larger losses also had higher mortality and, hence, larger weights, so that part of the difference in losses would have occurred even under the same relative increase in mortality. My results for the United States illustrate the size of this effect. Almost a quarter of its observed loss from 2019 to 2021 disappears in the counterfactual comparison. This does not call the conclusions of these studies into question, as the differences they report are much larger than the sensitivity effect alone could produce and inequalities in health and mortality are well-documented in the US. It shows, however, that the size of such differences can only be read as a difference in the development of mortality once all populations are evaluated with common weights. In epidemiology, crude death rates of populations with different age structures are not compared without age standardization. The same care is worthwhile in analyses of changes in life expectancy at birth.

In line with \citet{Cui2019} and \citet{Modig2020}, I therefore recommend examining changes in age-specific mortality rates alongside changes in life expectancy at birth, especially when populations with very different levels of mortality are compared. Counterfactual changes in life expectancy at birth and the decomposition into mortality and sensitivity effects offer a simple way to do so while keeping the intuitive scale of life expectancy.

\section{Conclusion}\label{sec:conclusion}

A change in life expectancy at birth combines two things, the change in age-specific mortality and the sensitivity of life expectancy to this change, which depends on the mortality schedule from which a population starts. The same relative change in mortality therefore produces larger gains or losses in populations with higher mortality and higher life disparity. Comparing changes in life expectancy at birth between such populations is, in this respect, comparable to comparing crude death rates between populations with different age structures.

Counterfactual changes in life expectancy at birth remove the second component by evaluating the changes in mortality of all populations with the same weights, and the decomposition into a mortality effect and a sensitivity effect shows how much of a difference in changes is due to each component and at which ages. Both measures are expressed in years (or months) of life expectancy, are straightforward to compute from standard life tables, and require no data beyond the death rates that are needed to calculate life expectancy itself.

My examples show that the sensitivity effect can be large enough to explain a substantial part, or even all, of a difference in changes in life expectancy, as for the sex differences between 1990 and 2019, but also that it does not always change the overall conclusions, as for the ranking of countries during the COVID-19 years 2020 and 2021. Whether the counterfactual comparison matters thus depends on how different the mortality schedules of the populations are and how similar their changes in mortality were. Before differences in changes in life expectancy at birth are interpreted as the result of health policies, health behaviour or living conditions, it should therefore be checked whether they reflect different developments of mortality or merely different starting points.

\section*{Acknowledgements}
I used Claude (Anthropic) during the preparation of this paper. It was used to write and debug the Python code and for producing the figures and tables. It was also used to draft parts of the text, in particular the description of the results, the abstract and the conclusion, and to check the numbers reported in the text against the output of the code. The research question, the methodological approach and the interpretation of the results are my own. I reviewed and revised the AI-generated code and text, verified the results, and take full responsibility for the content of this paper.

\FloatBarrier

\clearpage
\appendix
\section{Implementation}\label{app:implementation}

\subsection{From instantaneous to observed changes}\label{app:observed}

Eq.~\eqref{eq:e0dot_w} expresses the change in $e_0$ as a product of weights and changes in mortality, but only for an infinitesimal change. Data, in contrast, provide death rates for discrete years. A direct discrete analogue would combine the weights at the beginning of a period with the observed change in log death rates,
\begin{equation}
  \Delta e_0 \approx -\int_0^{\omega} w(x,t_1)\,\rho(x)\dd x,
  \qquad
  \rho(x) = \ln\mu(x,t_2)-\ln\mu(x,t_1).
  \label{eq:app_firstorder}
\end{equation}
Eq.~\eqref{eq:app_firstorder} is only a first-order approximation. Because $e_0$ is a non-linear function of the death rates, the weights themselves change while mortality changes, and the approximation leaves a residual that cannot be attributed to any age. In a decomposition, this residual would also prevent the mortality and sensitivity effects from adding up to the observed difference.

\paragraph{Integrating the instantaneous change.}
The exact change over a period is the integral of the instantaneous change,
\begin{equation}
  \Delta e_0 = \int_{t_1}^{t_2} \dot{e}_0(t)\dd t
  = -\int_{t_1}^{t_2}\int_0^{\omega} w(x,t)\,\rho(x,t)\dd x\dd t .
  \label{eq:app_integral}
\end{equation}
Evaluating Eq.~\eqref{eq:app_integral} requires the death rates at every point in time between $t_1$ and $t_2$, which are not observed. Following \citet{Horiuchi2008}, I therefore assume a path between the two observations. Their method applies to any function of several covariates: each covariate is assumed to change gradually between the two observations, and the change in the function is obtained by integrating its partial derivatives along this path. I use the log death rates, $\ln\mu(x,t)$, as covariates and assume that they change linearly in time, i.e., that mortality at each age changes at a constant relative rate:
\begin{equation}
  \ln\mu(x,t) = \ln\mu(x,t_1) + \frac{t-t_1}{t_2-t_1}\,\rho(x),
  \qquad t\in[t_1,t_2].
  \label{eq:app_path}
\end{equation}
Along this path, $\rho(x,t)=\rho(x)/(t_2-t_1)$ does not depend on $t$ and can be taken out of the time integral in Eq.~\eqref{eq:app_integral}:
\begin{equation}
  \Delta e_0 = -\int_0^{\omega} \bar{w}(x)\,\rho(x)\dd x,
  \qquad
  \bar{w}(x) = \frac{1}{t_2-t_1}\int_{t_1}^{t_2} w(x,t)\dd t .
  \label{eq:app_exact}
\end{equation}

\paragraph{Annual steps.}
Eq.~\eqref{eq:app_exact} could be applied directly to a long period, e.g., from 2010 to 2020. The total would still be exact, but the allocation to ages would rest on the assumption of a constant rate of change over ten years and would ignore the observed years in between. I therefore apply Eq.~\eqref{eq:app_exact} to each one-year interval, which gives Eq.~\eqref{eq:Delta_e0}, and sum the results:
\begin{equation}
  \Delta e_0 = \sum_{t=t_1}^{t_2-1}\Delta e_{0,t}
  = -\sum_{t=t_1}^{t_2-1}\int_0^{\omega}\bar{w}_t(x)\,\rho_t(x)\dd x .
  \label{eq:app_chain}
\end{equation}
The assumption of a constant rate of change then applies only within single years. The product of weight and change has to be formed within each year before summing over years. Averaging the weights and summing the changes over the whole period first would not reproduce $\Delta e_0$ unless the changes in mortality were the same in every year. The age contributions and the mortality and sensitivity effects are obtained in the same way, i.e., computed for each year and then summed.

\subsection{Life tables and weights}\label{app:lifetables}

For the year from $t$ to $t+1$, with $\rho_{x,t}=\ln m_x(t+1)-\ln m_x(t)$, the averaged weight $\bar{w}_{x,t}=\int_0^1 w_x(t+\tau)\dd\tau$ is computed from intermediate life tables with death rates
\begin{equation}
  \ln m_x(t+\tau) = \ln m_x(t) + \tau\,\rho_{x,t},
  \qquad \tau\in[0,1],
  \label{eq:app_intermediate}
\end{equation}
i.e., $m_x(t+\tau)=m_x(t)^{1-\tau}m_x(t+1)^{\tau}$. These intermediate life tables serve only to compute the weights. I calculated $K$ tables and average with Gauss--Legendre quadrature,
\begin{equation}
  \bar{w}_{x,t} \approx \sum_{k=1}^{K} v_k\, w_x(t+\tau_k),
  \label{eq:app_quadrature}
\end{equation}
where $\tau_k$ and $v_k$ are the Gauss--Legendre nodes and weights on $[0,1]$. With $K=10$, the age contributions $-\bar{w}_{x,t}\rho_{x,t}$ reproduce the observed $\Delta e_{0,t}$ to numerical precision; I use this identity as a check of the implementation. A single node at $\tau=1/2$, i.e., the life table with the geometric mean of the death rates of both years, already gives an error of third order in $\rho$. All integrals over age in Section~\ref{sec:methods} are replaced by sums over single years of age. Because $\rho_{x,t}$ is undefined when $m_x=0$, ages above 100 are aggregated into the open age interval.

\subsection{Counterfactual changes}\label{app:standardization}

\paragraph{Choice of the reference.}
As with age-standardized death rates, the reference should be a mortality schedule within the range of the populations compared. When two populations are compared, the average of their weights treats both symmetrically and links the counterfactual changes directly to the decomposition (Eq.~\eqref{eq:M_std}). When several populations are compared, the weights of the pooled life table avoid favouring any single population. When one population is followed over time, the weights of a fixed reference year are used. Otherwise later changes would appear smaller partly because life disparity, and hence the sensitivity of $e_0$, declines as life expectancy increases.

\subsection{Further decomposition of the sensitivity effect}\label{app:sensitivity}

The decomposition differs from the decomposition of crude death rates in two respects. First, the age structure of a population sums to one, so that the composition effect reflects differences in the age distribution only. The weights, in contrast, sum to life disparity, so that the sensitivity effect reflects differences in both the overall size of the weights and their distribution over age. With $\bar{e}^\dagger_{g,t}=\int_0^\omega \bar{w}_{g,t}(x)\dd x$ and normalized weights $\tilde{w}_{g,t}(x)=\bar{w}_{g,t}(x)/\bar{e}^\dagger_{g,t}$ for $g\in\{A,B\}$, the same identity as in Eq.~\eqref{eq:kitagawa} separates the two:
\begin{equation}
  \begin{aligned}
  S_t(x) = -\tfrac{1}{2}\big(\rho_{A,t}(x)+\rho_{B,t}(x)\big)
  \Big[&\underbrace{\tfrac{1}{2}\big(\tilde{w}_{A,t}(x)+\tilde{w}_{B,t}(x)\big)\big(\bar{e}^\dagger_{A,t}-\bar{e}^\dagger_{B,t}\big)}_{\text{level}} \\
  &+ \underbrace{\tfrac{1}{2}\big(\bar{e}^\dagger_{A,t}+\bar{e}^\dagger_{B,t}\big)\big(\tilde{w}_{A,t}(x)-\tilde{w}_{B,t}(x)\big)}_{\text{age pattern}}\Big].
  \end{aligned}
  \label{eq:S_split}
\end{equation}
The level part reflects that populations with a larger life disparity lose (or gain) more life expectancy for the same proportional change in mortality. The age-pattern part reflects whether the change in mortality occurs at ages that carry more weight in one population than in the other. For a change that is uniform across ages, the age-pattern part sums to zero over all ages, because $\int_0^\omega(\tilde{w}_{A,t}-\tilde{w}_{B,t})\dd x=0$, although it need not vanish at individual ages.

Second, the age structure of a population does not depend on its current mortality rates, whereas the weights are a function of mortality itself. The sensitivity effect is therefore a property of the mortality schedule and not of an external population structure.

\end{document}